# Microstructure and superconductivity of Ir-doped BaFe$_2$As$_2$ superconductor

X. L. Wang, H. Y. Shi, X.W. Yan, Y. C. Yuan, Z.-Y. Lu, X. Q. Wang, and T.-S. Zhao*

*Department of Physics, Renmin University of China, Beijing 100872, P. R. China*

Polycrystalline samples with nominal composition of Ba(Fe$_{1-x}$Ir$_x$)$_2$As$_2$ ($x$=0.10, 0.15, and 0.20) were investigated by means of X-ray diffraction (XRD), scanning electron microscopy (SEM), electrical resistivity, and magnetization measurements. XRD and SEM results showed that almost single phase samples were obtained. Bulk superconductivity with $T_C$~28 K was observed in the $x$=0.10 sample. $T_C$~28 K is the highest superconducting critical temperature among the reported data for electron-doped $A$Fe$_2$As$_2$-type ($A$=Ca, Sr, and Ba) superconductors. The upper critical field $H_{c2}(0)$ reaches as high as 65 T for the $x$=0.10 sample. The underlying physics is discussed in connection with Co-doping case.

* Corresponding author. Tel.: +86-10-6251-1523.

*E-mail address*: tszhao@ruc.edu.cn (T.-S. Zhao).

Since the recent discovery of superconductivity in the F-doped $R$FeAsO ($R$ = La, Ce, Pr, Nd, Sm, and Gd) and Th-doped GdFeAsO with the highest $T_C$ up to as high as 55-56 K [1-6], the FeAs-based superconductors have attracted much attention. Another family of oxygen-free FeAs-based compounds with the ThCr$_2$Si$_2$-type structure such as K or Na-doped $A$Fe$_2$As$_2$ ($A$ = Ca, Sr, Ba, and Eu) and Co-doped $A$Fe$_2$As$_2$ ($A$ = Sr, Ba) were found to be superconductors with the highest $T_C$ up to 38 K [7-13]. The electron-doped Ba(Fe, Co)$_2$As$_2$ system has been extensively studied and the temperature-composition phase diagram of this system has been determined by several groups [14-21]. The determined phase diagrams show a striking feature that there is a coexistence of antiferromagnetisim and superconductivity in an underdoped region. Very recently, a number of electron-doped Ba(Fe, $M$)$_2$As$_2$ ($M$ = Ni, Ru, Rh, and Pd) [22-25] and Sr(Fe, $M$)$_2$As$_2$ ($M$ = Ni, Ru, Rh, Pd, and Ir) [26-29] compounds was also found to exhibit superconductivity. Among the electron-doped superconductors, a higher $T_C$ was observed in Co-doped BaFe$_2$As$_2$ ($T_C$ ~ 25 K) [17, 18, 30, 31], SrFe$_{1.57}$Ir$_{0.43}$As$_2$ ($T_C$ ~ 24.2 K) [28], and BaFe$_{1.886}$Rh$_{0.114}$As$_2$ ($T_C$ ~ 24 K) [25], respectively. Here we report the observation of superconductivity in Ir-doped BaFe$_2$As$_2$ with $T_C$ up to 28 K, which is the highest value among the electron-doped $A$(Fe, $M$)$_2$As$_2$ superconductors reported so far.

Polycrystalline samples with nominal composition of Ba(Fe$_{1-x}$Ir$_x$)$_2$As$_2$ ($x$ = 0, 0.10, 0.15, and 0.20) were synthesized by solid state reaction method using BaAs and (Fe$_{1-x}$Ir$_x$)$_2$As as starting materials. The mixture of Ba grains and As grains was mixed, and that of Fe/Ir powders and As grains were carefully ground and then sealed in the quartz tubes under vacuum. The mixture of BaAs was slowly heated up to 500 °C for 10 h, 680 °C for 10 h, and then 730 °C for another 10 h, while the mixtures of (Fe$_{1-x}$Ir$_x$)$_2$As were slowly heated up to 500 °C for 10 h and then 800 °C for another 10 h. Stoichiometric amounts of BaAs and (Fe$_{1-x}$Ir$_x$)$_2$As powders were mixed, carefully ground and then pressed into pellets with a size of $\Phi 8 \times 1.5$ mm$^2$. The pellets were sintered in an evacuated quartz tube at 1100 °C for 20 h. In order to ensure sufficiently homogeneity of the samples, the Ir-containing pellets were reground and repressed, and finally sintered again at 1100 °C for 20 h. All the preparation processes for the mixing, grounding and pressing were carried out in a glove box under argon atmosphere with both H$_2$O and O$_2$ content $\leq$ 0.1 ppm.

Powder X-ray diffraction (XRD) using Cu $K_\alpha$ radiation at room temperature was used to identify phase structure of the samples. Figure 1(a) shows the powder XRD patterns for Ba(Fe$_{1-x}$Ir$_x$)$_2$As$_2$ ($x$ = 0, 0.10, 0.15, and 0.20) samples. One can see that the samples are almost of single phase, and main diffraction peaks can be well indexed with the tetragonal ThCr$_2$Si$_2$-type structure (space group $I4/mmm$). The determined lattice parameters of the $a$-axis and $c$-axis are listed in Table I. It can be seen from Table I that the $a$-axis increases and the $c$-axis shrinks slightly with increasing the Ir content $x$, respectively.

We have examined the microstructure and phase composition of the samples using a scanning electron microscope (SEM) (LEO 1450, Carl Zeiss SMT Ltd., Cambridge, UK) equipped with an energy dispersive X-ray (EDX) spectrometer. SEM micrographs show that, comparing with the undoped BaFe$_2$As$_2$, there is a gradual and considerable decrease in the grain size for the Ir-doped samples with increasing the Ir content. Figures 1(b) and 1(c) show the backscattered electron images for the samples with $x$ = 0.10 and 0.20, respectively. It can be seen that there exists a small fraction of white phase as an impurity phase in the samples. Our EDX analysis shows that the white phase is IrAs$_2$ phase. In fact, a small peak at 22.76$^o$ from AsIr$_2$ phase can be seen in Figure 1(a), and the intensity ratio between the (-111) diffraction peak of AsIr$_2$ phase and the (103) diffraction peak of Ba(Fe, Ir)$_2$As$_2$ phase is determined to be smaller than 5%. Thus, the actual Ir content in the samples will be slightly lower than the nominal Ir content. From the EDX analysis, the atomic ratio Ir/Fe was found to be 9.7/90.3, 14.0/86.0, and 18.8/81.2 (with an error within $\pm$ 5%) for the samples with $x$ = 0.10, 0.15, and 0.20, respectively. This means, for instance, the composition for the $x$ = 0.10 sample is close to Ba(Fe$_{0.903}$Ir$_{0.097}$)$_2$As$_2$. In addition, we notice that the existence of IrAs$_2$ phase may eventually inhibit the grain growth and result in a reduction of the grain size in the Ir-doped samples.

We have measured the electrical resistivity as a function of temperature for the samples by the standard four-probe method (PPMS, Quantum Design). Figure 2 shows the temperature dependence of the resistivity, $\rho$ ($T$), normalized to the resistivity at 300 K for Ba(Fe$_{1-x}$Ir$_x$)$_2$As$_2$ samples. One can see that superconducting (SC) transition occurs in the samples. The inset shows the enlarged $\rho$ ($T$) curves around $T_C$ for the samples. The determined onset SC transition and midpoint SC transition temperatures for the samples are listed in Table I. The highest SC transition temperature $T_C$ up to 28.2 K (onset) or 27.4 K (midpoint) with a transition width of 1.1 K is observed in the $x$ = 0.10 sample. To confirm the bulk nature of the superconductivity in the samples, the magnetization as a function of temperature was measured by means of a vibrating sample magnetometer (PPMS, Quantum Design). Figure 3(a) shows the temperature dependence of the magnetization for the samples measured under conditions of zero field cooling (ZFC) and field cooling (FC) at $H$ = 20 Oe. The samples show strong diamagnetic signals, indicating bulk superconductivity in Ba(Fe$_{1-x}$Ir$_x$)$_2$As$_2$ samples. The onset SC transition temperature determined from the ZFC curve is 27.3 K, 27.0 K, and 24.7 K for the samples $x$ = 0.10, 0.15, and 0.20, respectively, which are in agreement with the $T_C^{mid}$ values determined from $\rho$ ($T$). $T_C$ ~ 28 K is the highest SC transition temperature among the electron-doped $A$Fe$_2$As$_2$-type ($A$ = Ca, Sr, and Ba) superconductors reported so far [12-32].

Figure 3(b) shows the magnetoresistive SC transition under different applied fields for the $x$ = 0.10 sample. One can see that $T_C$ decreases and the transition width becomes broadening slightly with increasing magnetic fields. The inset of Fig. 3(b) plots the upper critical field $H_{c2}(T)$ as a function of temperature obtained from a determination of the midpoint of the resistive transition. The slope $-\mu_0(dH_{c2} / dT)|_{T_c}$ is found to be 3.46 T/K. The value for the upper critical field $H_{c2}(0)$ can be estimated by using the Werthamer–Helfand–Hohenberg (WHH) formula, $H_{c2}(0) = -0.69T_c (dH_{c2} / dT)|_{T_c}$ [33]. Thus obtained value for $\mu_0 H_{c2}(0)$ is 65.4 T for the $x$ = 0.10 sample. This value for $H_{c2}(0)$ is very high and comparable to that of (Ba, K)Fe$_2$As$_2$ with $T_C$ ~ 28 K [34] and Ba(Fe, Co)$_2$As$_2$ with $T_C$ ~ 22 K [35-36].

The ground state of an iron pnictide is in a collinear antiferromagnetic order. The superconductivity takes place after the doping or the high pressure suppresses the long-range magnetic order. It is very likely that the superconductivity pairing is mediated by the antiferromagnetic spin fluctuations. An iron pnictide has the hole and electron Fermi pockets at Γ point and M point, respectively, with a wavevector of (π, π) connected. And a number of experiments have suggested a spin singlet pairing between the hole Fermi pocket and the electron Fermi pocket [37-39]. We have carried out density functional theory (DFT) electronic structure calculations for Co/Ir-doped BaFe$_2$As$_2$. The doping effect upon the electronic structures was studied by using virtual crystal calculations. We find that the Fermi surfaces have very similar changes after Co-doping and Ir-doping with $x \leq 0.10$. This suggests there would be the similar superconductivity behavior. For the doping level $x$ > 0.10, the hole pockets at Γ point substantially shrink, where the Co-doping makes more serious damage than Ir-doping, as shown in Figures 4(a) and 4(b) for $x$ = 0.15. In experiment, it was found that $T_C$ for $x$ = 0.15 Co-doping has been dramatically reduced to less than 10 K [14-21]. In contrast, $T_C$ for $x = 0.15$ Ir-doping remains as high as 27 K (see Table I). The Co/Ir-doping induced Fermi surface topological changes are thus in consistency with the doping induced superconductivity behavior.

On the other hand, the electronic structure calculations for the Ir-doping show that there exists a strong hybridization between the Ir $5d$ and As $4p$ orbitals owing to the nature of much extended $5d$ orbitals. This strong hybridization in the Ir-doped BaFe$_2$As$_2$ may be relevant to the enhancement of $T_C$ as compared to Co and Rh-doped BaFe$_2$As$_2$ [17, 18, 25, 30, 31].

Furthermore, a series of experiments have shown that the maximum value of $T_C$ is higher when the bond angle of As-Fe-As is closer to 109.47° (regular FeAs$_4$ tetrahedron angle) [40, 41]. Figure 4(c) shows that the lattice parameter ratio of $c/c_0$ decreases more rapidly by Co-doping than Ir-doping. A decrease in the lattice parameter $c$ means an increase in the bond angle of As-Fe-As. Thus it is expected that, for a given doping level $x$, the bond angle of As-Fe-As for Ir-doping is lower than that for Co-doping. This is a possible reason why Ir-doping induced maximum $T_C$ is higher than Co-doping induced one.

In conclusion, we have studied the microstructure and superconductivity of polycrystalline Ba(Fe$_{1-x}$Ir$_x$)$_2$As$_2$ samples. XRD and SEM results showed that almost single phase samples were obtained, except for a small amount of IrAs$_2$ phase as an impurity phase. The existence of IrAs$_2$ phase may effectively reduce the grain size in the samples. Bulk superconductivity with $T_C \sim 28$ K was observed in the $x = 0.10$ sample. $T_C \sim 28$ K is the highest superconducting critical temperature among the reported data for the electron-doped $A$(Fe, M)$_2$As$_2$-type ($A$ = Ca, Sr, and Ba) superconductors. The upper critical field $H_{c2}(0)$ was estimated to be as high as 65 T for the $x = 0.10$ sample. The doping effect upon the Fermi surface topology and the bond angle of As-Fe-As is investigated in compare with Co-doping case.

The authors would like to thank G. F. Chen for helpful discussions and F. E. Cui for the assistance with SEM measurements. This work was supported by the National Basic Research Program of China (Contract No. 2007CB925001) and by the NSFC (Grant No. 20673133, 10874244, and 10974254)..

Table I. Lattice parameters of the *a*-axis and *c*-axis, onset SC transition and midpoint SC transition temperatures for Ba(Fe$_{1-x}$Ir$_x$)$_2$As$_2$ samples.

| x | a (Å) | c (Å) | $T_C^{onset}$ (K) | $T_C^{mid}$ (K) |
|---|---|---|---|---|
| 0 | 3.961 | 13.017 | --- | --- |
| 0.10 | 3.975 | 12.997 | 28.2 | 27.4 |
| 0.15 | 3.989 | 12.985 | 28.2 | 27.0 |
| 0.20 | 3.987 | 12.981 | 26.4 | 24.5 |

**Figure Captions**

Figure 1. (color online) Powder x-ray diffraction patterns for Ba(Fe$_{1-x}$Ir$_x$)$_2$As$_2$ samples with $x$ = 0, 0.10, 0.15, and 0.20. Backscattered electron images for the samples with $x$ = 0.10 (b) and 0.20 (c). The white phase is the IrAs$_2$ phase.

Figure 2. (color online) Temperature dependence of resistivity normalized to $\rho$ (300 K) for Ba(Fe$_{1-x}$Ir$_x$)$_2$As$_2$. The curves are vertically offset and separated by 0.2 units for clarity. Inset: enlarged temperature dependence of the resistivity around $T_C$.

Figure 3. (color online) (a) Temperature dependence of magnetization for Ba(Fe$_{1-x}$Ir$_x$)$_2$As$_2$ samples. The data was measured under conditions of zero field cooling (ZFC) and field cooling (FC) at $H$ = 20 Oe. (b) Temperature dependence of the resistivity, $\rho$ (T), for the $x$ = 0.10 sample under different applied fields from 0 to 10 T. Inset: the upper critical filed $H_{c2}$ (T) as a function of temperature.

Figure 4. (color online) Calculated Fermi surfaces of Ba(Fe$_{1-x}$M$_x$)$_2$As$_2$ with $x$ = 0.15. (a) $M$ = Co, (b) $M$ = Ir. (c) Lattice parameter ratio of $c/c_0$ as a function of x for Ba(Fe$_{1-x}$M$_x$)$_2$As$_2$ ($M$ = Co and Ir). Data for Ba(Fe$_{1-x}$Co$_x$)$_2$As$_2$ is taken from Ref. [14].

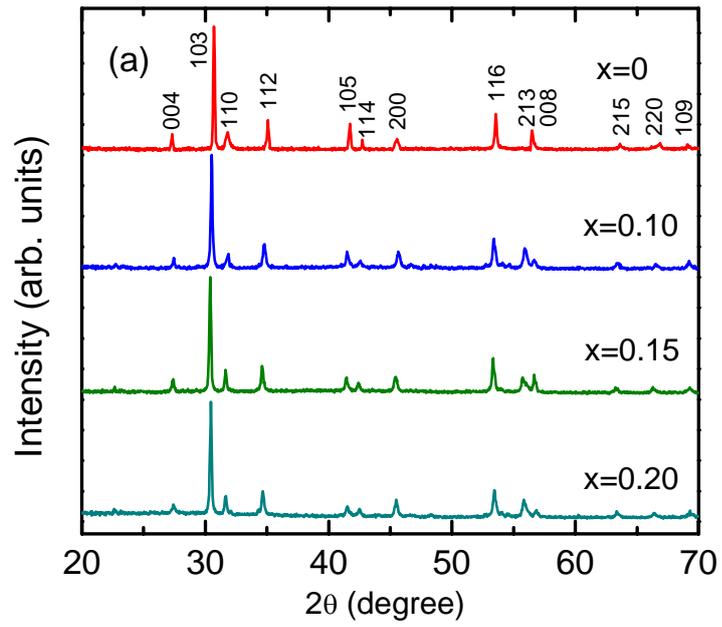

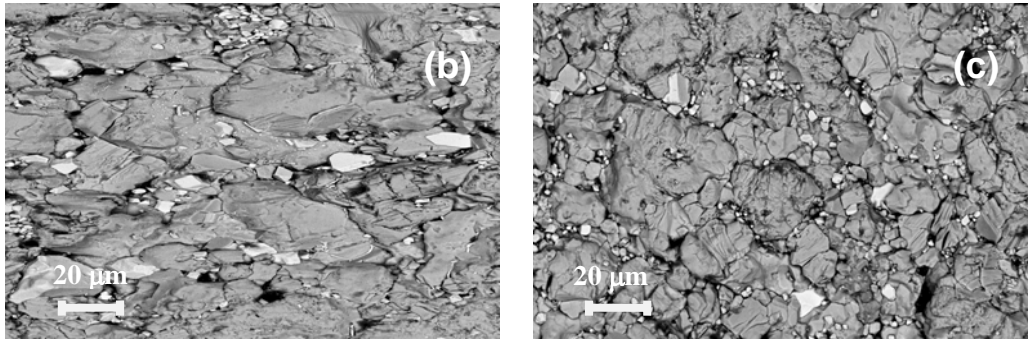

Figure 1

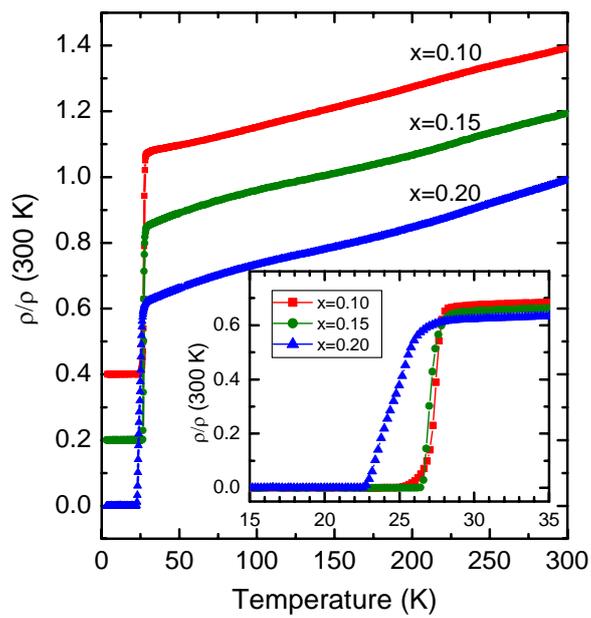

Figure 2

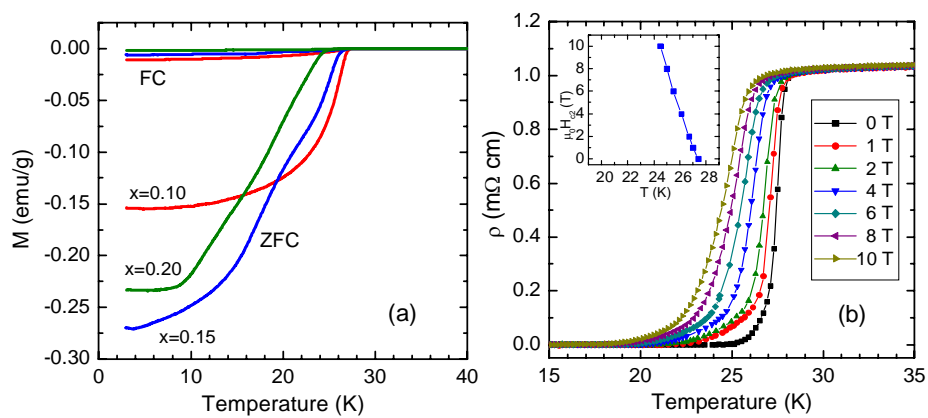

Figure 3

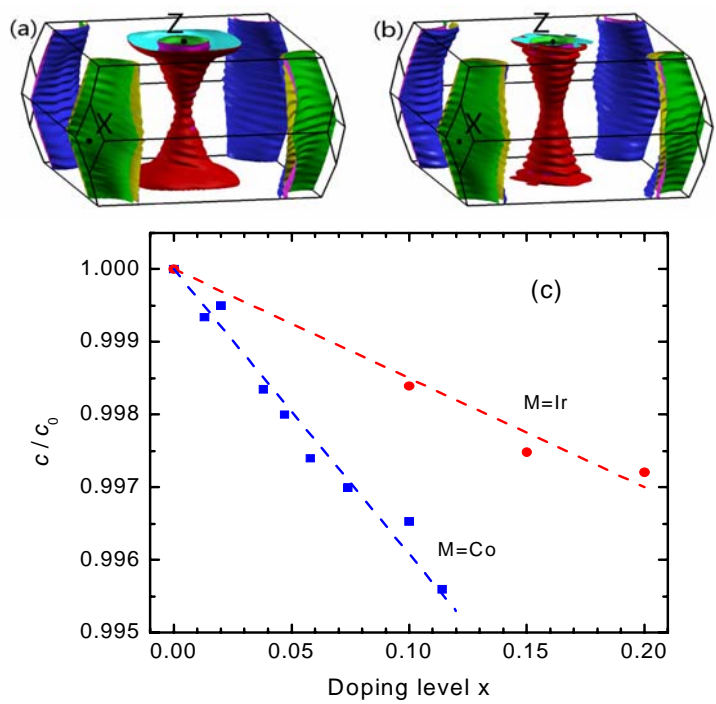

Figure 4